\documentclass[amsmath,amssymb,showkeys,superscriptaddress,aps,prd,10pt,onecolumn]{revtex4-2}
\usepackage{graphicx}
\usepackage[utf8]{inputenc}
\usepackage[caption=false]{subfig}
\usepackage{multirow}
\usepackage{appendix}
\usepackage{graphicx,epstopdf}

\usepackage{colortbl}
\usepackage{xcolor}
\usepackage[colorlinks=true, urlcolor=blue, linkcolor=blue, citecolor=blue]{hyperref}
\usepackage{float}
\usepackage{tikz}
\usetikzlibrary{arrows.meta}

\begin{document}

\title{Revisiting light-flavor diquarks in the inverse matrix method of QCD sum rules}
\author{Halil Mutuk}%
\email[]{hmutuk@omu.edu.tr}
\affiliation{Department of Physics, Faculty of Sciences, Ondokuz Mayis University, 55200, Samsun, Türkiye}

 
\begin{abstract}
This study reexamines the spectroscopic parameters of light-flavor diquarks within the framework of quantum chromodynamics sum rules (QCDSR) using the inverse matrix method. Conventional QCDSR analyses are based on assumptions such as quark-hadron duality and continuum models, which introduce a degree of systematic uncertainty. The inverse matrix method circumvents these assumptions by reformulating the problem as an inverse integral equation and expanding the unknown spectral density using orthogonal Laguerre polynomials. This method allows for a direct determination of spectral densities, thereby enhancing the precision of predictions regarding resonance masses and decay constants. By employing this methodology with regard to light-flavor diquarks ($sq$ and $ud$), it is possible to extract the associated masses and decay constants. The results indicate that the masses of diquarks with quantum numbers $J^P = 0^+$ and $J^P = 0^-$ are nearly degenerate. We compare our results regarding masses and decay constants with those of other theoretical predictions, which could prove a useful complementary tool in interpretation. Our results are consistent with those in the literature and can be shown as evidence for the consistency of the method. The results achieved in this study highlight the potential of the inverse matrix method as a robust tool for exploring nonperturbative QCD phenomena and elucidating the internal structure of exotic hadronic systems.
\end{abstract}

\maketitle

\section{Prologue}\label{introduction}

One of the primary objectives in hadronic physics is to gain a comprehensive understanding of the hadron spectrum as it pertains to quantum chromodynamics (QCD), the fundamental theory governing the strong interaction. In the field of quantum mechanics, stationary states that arise in bound state problems are eigenstates of an appropriate Hamiltonian. In contrast, bound states in quantum chromodynamics (QCD) are considerably more complex entities. In QCD, the excitation energy is enough to produce multiple constituents. In hadrons, a typical excitation energy is a few MeV due to QCD. This is adequate to create one or more light quark-antiquark pairs. In QCD, the property of asymptotic freedom enables the use of perturbation theory to calculate hard processes, which are characterized by short distances. Conversely, bound states (or resonances) of quarks and gluons emerge due to the significant effects of strong coupling at large distances, which cannot be addressed within the framework of perturbation theory. Given its intrinsic nature as a strongly interacting many-body theory, coupled with its strong coupling and confinement characteristics, solving QCD at low energies relevant to hadronic binding is a formidable challenge. This low-energy regime of QCD is an important phenomena which needs to be handled in order to gain more insight in the strong interactions.

The QCD sum rules (QCDSR) method deals with this nonperturbative regime of QCD and is an effective technique for elucidating the profound interrelationship between hadron phenomenology and the structure of the QCD vacuum through the analysis of a few condensate parameters, which are expectation values of QCD local operators \cite{Shifman:1978bx,Shifman:1978by}. This nonperturbative method has been successfully applied to a variety of problems in order to gain a field-theoretical understanding of the structure of hadrons. The QCDSR method is based on the evaluation of an appropriate correlation function within the deep Euclidean region through the application of the operator product expansion (OPE) on the one hand, and its phenomenological assessment through the examination of physical hadronic states on the other hand.  

In order to study the properties of hadrons, this method represents hadrons via their interpolating currents. A QCDSR calculation is comprised of three principal components: Firstly, an approximation of the correlation function is made in terms of QCD degrees of freedom through an OPE. Secondly, a description of the same correlator is provided in terms of the physical intermediate states. The aforementioned states are then represented through a dispersion relation that incorporates a straightforward ansatz for the spectral density. Finally, a methodology is utilized for matching these two descriptions and extracting the parameters of the spectral function that characterize the hadronic state of interest. In practical terms, a suitably chosen correlation function is calculated in two different frameworks: the hadronic framework and OPE framework. These two frameworks are then matched to each other using their spectral representation. This correspondence between theoretical and practical perspectives is often met with skepticism by those who are not practitioners, see Refs. \cite{Cohen:1994wm,Nielsen:2009uh,Albuquerque:2018jkn} for the formal answers. This matching results from the principle of duality, which establishes a connection between a description based on the fundamental constituents of matter, namely hadrons, and one based on the underlying dynamics of quarks and gluons, and is the first assumption of QCDSR formalism. The second key assumption is the utilization of the continuum model to eliminate the contamination of excited states from the hadron correlator under examination.  Borel transformation can be applied to suppress contribution of continuum and higher states on hadron side. In addition to QCD input parameters, there are two further parameters arise in the QCDSR approach: continuum threshold parameter and Borel parameter. A sum rule window should be obtained under some constraints to ensure that extracted properties from sum rules should not depend heavily on these parameters. The QCDSR technique employs certain phenomenological inputs, which constrains the precision of the method to a range of approximately $10 \% - 20 \%$ \cite{Narison:1989aq,Narison:2002woh}.

New approaches have recently been proposed to improve the QCDSR method. One of these methods is the inverse matrix method. This method may be considered one of the methodologies for addressing inverse problems. The unknown spectral density is expanded using Laguerre polynomials, and a matrix equation is established by equating the coefficients of $1/(q^2)^m$  on both sides of the sum rule. The matrix equation should be solved in order to obtain an approximate solution for the spectral density. This method does not require assumptions like quark-hadron duality and accurately reproduces the ground state peak.

In this work we reevaluate the spectroscopic parameters of light-flavor diquark states using inverse matrix method of QCDSR formalism. To the best of our knowledge, there is no work about diquarks within the framework of this method.

This paper is structured as follows: in Sec. \ref{formalism}, we briefly review the general aspects of the conventional QCDSR technique. In Sec. \ref{inverseQCD}, we introduce inverse problem of QCDSR formalism. The numerical results are presented, accompanied by a discussion of their relevance in Sec. \ref{numerical}. This work ends with a summary in Sec. \ref{final}.

\section{QCD sum rule method} \label{formalism}

To study hadron properties, this method represents hadrons via their interpolating currents. A correlation function is calculated in a hadronic and an OPE framework. Then, they are matched using their spectral representations.

In QCDSR, distinguishing long-distance and short-distance quark-gluon interactions is crucial. In the short-distance regime, where the energy scale is low, the behavior of quarks can be described as that of free particles. In this regime, perturbative QCD provides a suitable theoretical tool. Conversely, in the long-distance region, quark-gluon interactions are the dominant force, and quarks can be described in terms of condensates.

A QCDSR calculation starts with the correlation function
\begin{eqnarray}
\Pi(q^2) &=& i \int d^4x e^{iqx} \langle 0 | T  \bigl\{ J(x) J^\dagger (0) \bigr\} |
0 \rangle, \label{corrfunc}
\end{eqnarray}
where $J(x)$ is the interplating current and has the same quantum numbers with corresponding state $H$ and $T$ is the time ordering operator. The coupling of $J(x)$ to $H$  is written as:
\begin{eqnarray}
f_H  \equiv  \langle 0 | J(0) | H \rangle.
\end{eqnarray}

The correlation function is expressed as a dispersion relation at the hadronic level:

\begin{equation}
\Pi(q^2)={\frac{1}{\pi}}\int^\infty_{s_0}\frac{{\rm Im} \Pi(s)}{s-q^2-i\varepsilon}ds \, ,
\end{equation}
where the lower limit of integration is the physical threshold.  The imaginary part of the correlation function is the spectral density:
\begin{eqnarray}
\rho_{\rm phen}(s) \equiv {1 \over \pi}{\rm Im} \Pi(s) = \sum_n\delta(s-M^2_n)\langle 0| J |n\rangle\langle n| J^\dagger
|0\rangle \, .
\end{eqnarray}

A parametrization of single pole dominance and continuum contribution is usually adopted for the ground state H
\begin{eqnarray}
\rho_{\rm phen}(s) =  f^2_H \delta(s-M^2_H) + \rm{excited\,\,states}\, .
\end{eqnarray}

At the quark-gluonic level $\Pi(q^2)$ can be evaluated within the OPE framework and evaluates the spectral density $\rho_{\text{OPE}}(s)$ up to a specified order.  Two representations of correlation function then can be equated as:

\begin{equation}
\int_0^\infty ds \frac{\rho_{\text{phen}}(s)}{s-q^2} + \mbox{polynomials in $q^2$}= \int_0^\infty ds \frac{\rho_{\text{OPE}}(s)}{s-q^2} + \mbox{polynomials in $q^2$}.
\end{equation}

Borel transformation makes the transformation 
\begin{equation}
\mathcal{B}_{M_B^2}\Pi(q^2) =
\int^\infty_{s_0} e^{-s/M_B^2} \rho(s) ds,
\end{equation}
where the main effect of the Borel transformation is to get rid of polynomials while at the same time makes the transformation:

\begin{equation}
\frac{1}{s-q^2} \rightarrow e^{-\frac{s}{M^2}}.
\end{equation}

If the contribution of the continuum states to the OPE spectral density can be approximated with sufficient accuracy above a specified threshold value $s_0$, then one can derive a sum rule relation:

\begin{equation}
f^2_H e^{-M_H^2/M_B^2} = \int^{s_0}_0 e^{-s/M_B^2}\rho_{\text{OPE}}(s)ds
\label{bortrans} \, .
\end{equation}

The QCD sum rules can be derived by carrying out an analysis of this final expression. The sum rules for mass can be derived by differentiating both sides of the equation with respect to $1/M_B^2$:

\begin{equation}
M^2_H = \frac{\int^{s_0}_0 e^{-s/M_B^2} s \rho_{\text{OPE}}(s) ds} {\int^{s_0}_0
e^{-s/M_B^2} \rho(s) ds} \label{masssumrule}
\end{equation}

In the same way, the pole residue (decay constant) can be obtained via

\begin{equation}
f^2_H = e^{M_H^2/M_B^2} \int^{s_0}_0 e^{-s/M_B^2}\rho_{\text{OPE}}(s)ds
\label{poleresidue} \, .
\end{equation}

The extraction of physical properties (mass, decay constant, witdh, etc.) from the obtained sum rules in standard QCDSR analysis follows three stages: (i) The assumption of continuum threshold $s_0$ which starts, based on the experience, near from the first excited states. In conventional mesons and baryons, the parameters of the first excited states are either known through experimental observation or have reached a theoretical consensus. In the case of exotic states, there is often a lack of available information regarding both excited states and ground states. (ii) A suitable working window (also referred to as Borel window) is defined. The lower and upper limits of this interval are determined through the application of pole dominance and OPE convergence. (iii) In this working window, the stability of the sum rules is investigated to determine a suitable continuum threshold $s_0$ and Borel parameter $M_B^2$, and physical observables such as mass and decay constant are obtained.

\section{Inverse problem of QCDSR formalism}\label{inverseQCD} 

In this section, we introduce inverse matrix method of QCDSR formalism. We will apply this method on light diquarks. Similar to quarkonium $q \bar{q}$ states, when two quarks come together, they can form bound states $qq$ due to the attractive force between them. This attractive force is a result of the spins and color charges of the two quarks being different, or more precisely, their spins and color charges being antisymmetric. The strongest evidence for these correlations is found when they involve the light quarks \cite{Alexandrou:2006cq}. Actually, the concept of diquarks originated at the outset of the development of the quark model \cite{Gell-Mann:1964ewy,Zweig:1964jf}. The reappearance of interest in diquarks in the context of hadronic spectroscopy can be traced back to 2004 \cite{Jaffe:2003sg,Jaffe:2004ph,Wilczek:2004im,Selem:2006nd}. These works identified the spin-0 flavor-antisymmetric diquarks and designated them as ``good diquarks", in contrast to flavor-symmetric diquark states, which were termed ``bad diquarks". Good diquarks are in the color-antitriplet representation. Nowadays, it is believed that some exotic states are composed of diquarks (see a recent review and references therein \cite{Barabanov:2020jvn}).

Within the standard analysis, QCDSR formalism predicts many physical properties that agree well with the experimentally observed phenomena. In addition to this, there are alternative ways applied to improve this formalism. In Ref. \cite{Leinweber:1995fn}, a satisfactory answer to predictive ability of QCDSR formalism was obtained by utilizing the Monte-Carlo method. Semileptonic B meson decay is investigated via QCD light-cone sum rule by invoking a method that bypasses the semi-global quark-hadron duality approximation, which typically introduces an unknown and potentially significant systematic error in the prediction of form factors \cite{Carvunis:2024koh} . In Refs. \cite{Li:2020xrz,Li:2020ejs,Xiong:2022uwj}, QCDSR formalism is handled as an inverse problem in which the objective is to utilize the calculated results at the QCD level as input to solve the hadron spectral density in a dispersion integral. In Refs. \cite{Li:2021gsx,Li:2022qul,Li:2022jxc,Li:2023dqi,Li:2023yay,Li:2023ncg,Li:2023fim,Li:2024awx,Li:2024awx,Li:2024xnl,Zhao:2024drr,Li:2024fko}, the inverse matrix method is applied to QCDSR formalism. In this method, the unknown spectral density embedded within the dispersion integral is expanded using generalized Laguerre polynomials. Then, equating the coefficients of $1/(q^2)^m$  on both sides establishes a matrix equation.

In Refs. \cite{Dosch:1988hu,Jamin:1989hh}, the flavor $(sq)$ diquark current was taken to be 
\begin{equation}
j_i(x)=\epsilon_{ijk}s^T_j(x)COq_k(x). \label{dicurrent}
\end{equation}
Here, the indices $i,~j,~k$ refer to colors, the charge conjugation matrix is represented by the symbol $C$, and the Lorentz structures $O$ are given by the expressions $O = \gamma_5,~1$. These correspond to the quantum numbers $J^{P} = 0^+, \text{and} \ 0^-$, respectively. Inserting Eq. (\ref{dicurrent}) into the Eq. (\ref{corrfunc}) yields the correlation function for $J^P=0^{\pm}$ quantum numbers as \cite{Dosch:1988hu,Jamin:1989hh}:

\begin{eqnarray}
\Pi(q)&=&{-\frac{3}{4}\pi^2}\left(1-2{m^2_s\over q^2}+{17\over
6}{\alpha_s\over \pi}-{1\over 2}{\alpha_s\over \pi}\ln{-q^2\over
\mu^2}\right) q^2\ln{-q^2\over \mu^2}+ (\pm 2m_s-m_q){\langle \bar
qq\rangle\over q^2}-(\pm 2m_q-m_s){\langle \bar ss\rangle\over
q^2}\\\nonumber &-& {1\over 8}\langle {\alpha_s\over \pi}G^2\rangle{1\over
q^2} \pm 8\pi\kappa\alpha_s{\langle \bar
qq\rangle\langle\bar ss\rangle\over q^4}-{16\pi\over
27}\kappa\alpha_s{\langle \bar qq\rangle^2+\langle\bar
ss\rangle^2\over q^4},
\end{eqnarray}
where $\langle\bar qq\rangle$, $\langle\bar ss\rangle$, $\langle {\alpha_s\over\pi} G^2\rangle$ are quark and gluon condensates, $\kappa$ denotes the deviations from the factorization approximation of condensates of higher dimensions and $\mu$ being a normalization scale. Replacing $s$ with $u$  gives the result of $(qq)$ diquark.

The QCDSR is an effective tool for examining the nonperturbative elements of QCD. Conventional methods often use Borel transformations and stability analysis based on the quark-hadron duality assumption. However, a novel approach treats QCDSR as an inverse problem, which allows for a more direct and potentially more accurate determination of spectral densities without these assumptions.

In the inverse problem approach to QCDSR, the spectral density on the hadron side is not posited to adhere to the principles of quark-hadron duality. In contrast, both the resonance and the continuum contributions to the spectral density are determined with the help of the perturbative input from the quark side. This approach eliminates the necessity for a continuum threshold and Borel transformation, thereby simplifying the analysis and potentially increasing accuracy.

\begin{figure}[ht]
\centering
\includegraphics[width=10cm]{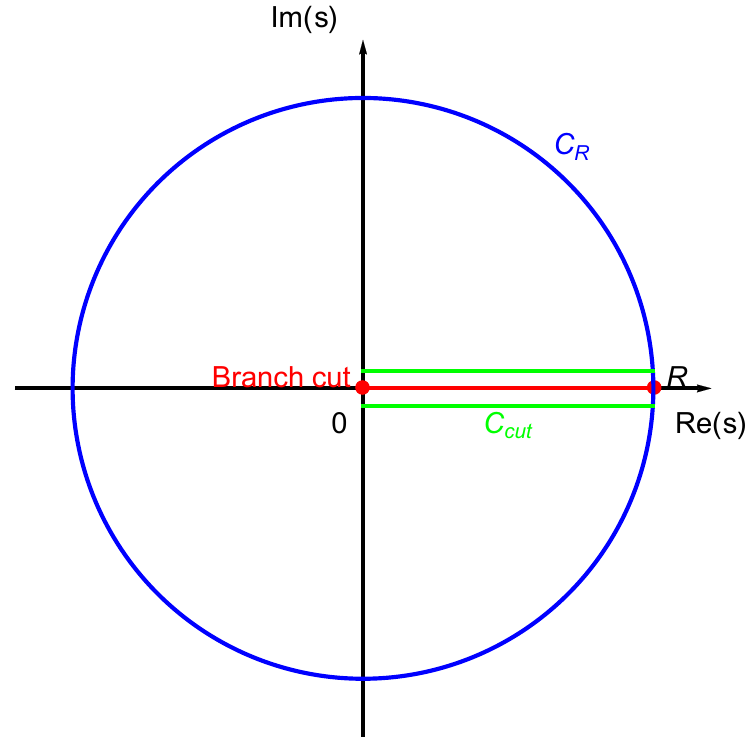}
\caption{Contour $C$ in the complex $s$-plane, consisting of the circle $C_R$ with radius $R$ and the lines $C_{\text{cut}}$ enclosing the branch cut along the positive real axis.}
\label{contour}
\end{figure}

The correlation function $\Pi(q^2)$, defined in Eq. (\ref{corrfunc}), is analytic in the complex plane except along the positive real axis, where it may exhibit poles and branch cuts. This analytic structure allows us to express $\Pi(q^2)$ via a contour integral in the complex $s$-plane:
\begin{equation}
\Pi(q^2) = \frac{1}{2\pi i} \oint_C \frac{\Pi(s)}{s - q^2} ds, \label{corrfunc1}
\end{equation}
where the contour $C$, depicted in Fig. \ref{contour}, consists of a circle $C_R$ with a large radius $R$, and lines $C_{\text{cut}}$ running above and below the positive real axis, enclosing the branch cut. This representation separates the nonperturbative and perturbative contributions to $\Pi(q^2)$. 

The contour integral in Eq. (\ref{corrfunc}) can be decomposed into contributions from the circle $C_R$ and the cut $C_{\text{cut}}$ as
\begin{eqnarray}
\Pi(q^2) = \frac{1}{2\pi i} \left( \oint_{C_R} \frac{\Pi(s)}{s - q^2} ds + \oint_{C_{\text{cut}}} \frac{\Pi(s)}{s - q^2} ds \right). \label{corrfunc2}
\end{eqnarray}
Here, this separation clarifies how the hadronic spectral function is extracted while maintaining consistency with the analytic structure of the correlation function. It ensures that the sum rules correctly relate QCD calculations to hadronic observables without ambiguities in contour integration.

The perturbative calculation of $\Pi(s)$ is valid for large $s$, away from the physical poles. Thus, the integral over $C_R$ can be evaluated using the perturbative expression $\Pi^{\text{pert}}(s)$. On the other hand, the integral over $C_{\text{cut}}$ captures the nonperturbative dynamics near the branch cut, which we express in terms of the imaginary part of $\Pi(s)$:
\begin{equation}
\frac{1}{2\pi i} \oint_C \frac{\Pi(s)}{s - q^2} ds = \frac{1}{\pi} \int_0^R \frac{\text{Im}\, \Pi(s)}{s - q^2} ds + \frac{1}{2\pi i} \oint_{C_R} \frac{\Pi^{\text{pert}}(s)}{s - q^2} ds,
\end{equation}
where the first term on the right-hand side represents the nonperturbative contribution from the cut $C_{\text{cut}}$, where $\text{Im}\, \Pi(s)$ encodes the spectral density of hadronic states and the second term represents the perturbative contribution from the circle $C_R$, where $\Pi^{\text{pert}}(s)$ is calculated using OPE techniques. The decomposition highlights the interplay between perturbative and nonperturbative effects in QCD: (i) the perturbative term $\Pi^{\text{pert}}(s)$ dominates at large $s$, where asymptotic freedom ensures reliable calculations and (ii) the nonperturbative term $\text{Im}\, \Pi(s)$ encodes the hadronic spectral density, which is treated as an unknown quantity in the context of the inverse problem. This term is crucial for extracting physical quantities such as masses and decay constants.

The OPE expansion of the left hand side of Eq. (\ref{corrfunc1}) reads as
\begin{eqnarray}
\Pi^{\rm OPE}(q^2)&=&\frac{1}{2\pi i}\oint \frac{\Pi^{\rm pert}(s)}{s-q^2}ds+
(\pm 2m_s-m_q){\langle \bar
qq\rangle\over q^2}-(\pm 2m_q-m_s){\langle \bar ss\rangle\over
q^2}- {1\over 8}\langle {\alpha_s\over \pi}G^2\rangle{1\over
q^2} \nonumber \\  &\pm & 8\pi\kappa\alpha_s{\langle \bar
qq\rangle\langle\bar ss\rangle\over (q^2)^2}-{16\pi\over
27}\kappa\alpha_s{\langle \bar qq\rangle^2+\langle\bar
ss\rangle^2\over (q^2)^2},\label{corrope}
\end{eqnarray}
where $\Pi^{\rm pert}(s)$ has been expressed as an integral over the same contour. The perturbative calculation of $\Pi(s)$ is reliable for s distant from the physical region, and $\Pi^{\rm pert}(s)$  has only the perturbative QCD contribution. The integrand along the large circle C with R has been replaced by $\Pi^{\rm pert}(s)$.

Equating hadron side to quark side will give to sum rules of QCD. Matching Eq. (\ref{corrfunc2}) on the quark side and Eq. (\ref{corrope}) on the hadron side will lead to the sum rule as
\begin{eqnarray}
\frac{1}{\pi}\int_{0}^R \frac{\text{Im}\Pi(s)}{s-q^2} ds &=& \frac{1}{\pi}\int \frac{\Pi^{\rm pert}(s)}{s-q^2}ds + (\pm 2m_s-m_q){\langle \bar
qq\rangle\over q^2}-(\pm 2m_q-m_s){\langle \bar ss\rangle\over
q^2}- {1\over 8}\langle {\alpha_s\over \pi}G^2\rangle{1\over
q^2} \nonumber \\  &\pm & 8\pi\kappa\alpha_s{\langle \bar
qq\rangle\langle\bar ss\rangle\over (q^2)^2}-{16\pi\over
27}\kappa\alpha_s{\langle \bar qq\rangle^2+\langle\bar
ss\rangle^2\over (q^2)^2}, \label{matching}
\end{eqnarray}
where in regions away from the poles, the perturbation function $\Pi^{\text{pert}}(s)$ is cancelled out, leaving only the perturbative spectral density
\begin{equation}
 \frac{1}{\pi} \Pi^{\text{pert}}(s)= {-\frac{3}{4}\pi^2}\left(1-2{m^2_s\over q^2}+{17\over
6}{\alpha_s\over \pi}-{1\over 2}{\alpha_s\over \pi}\ln{-q^2\over
\mu^2}\right) q^2\ln{-q^2\over \mu^2},  \label{rhospec}
\end{equation}
remains along the branch cut. In conventional QCDSR applications, Borel transformations are employed to mitigate the impact of uncertain continuum contributions on the hadron side and to enhance the perturbative expansion on the quark side. Alternatively, following Refs. \cite{Li:2021gsx,Zhao:2024drr}, a new definition for the generic spectral density $\rho(s) \equiv \frac{1}{\pi} \text{Im}\Pi(s)$ could be done as
\begin{equation}
\rho(s)=\Delta \rho(s, \Lambda) +  \frac{1}{\pi} \Pi^{\text{pert}}(s) (1- e^{-s/\Lambda}), \label{newrhos}
\end{equation}
where $\Delta \rho(s, \Lambda)$ is called ``subtracted spectral density" which can be obtained readily from Eq. (\ref{newrhos}) as
\begin{equation}
\Delta \rho(s, \Lambda)= \rho(s)-\frac{1}{\pi} \Pi^{\text{pert}}(s) (1- e^{-s/\Lambda}).
\end{equation}
Here, $\Lambda$ describes the transition from $\text{Im}\Pi(s)$ to to $\text{Im}\Pi^{\text{pert}}(s)$, which is a function of the perturbative parameter. The function $(1- e^{s/\Lambda})$ behaves like $s$: it decreases at small $s$ and approaches unity at large $s$. In this case $\Delta \rho(s, \Lambda)$ treats to be $\rho(s) \sim s$ in the limit of $s \to 0$, and decreases rapidly as $s > \Lambda$. The R in Eq. (\ref{matching})  can be extended to infinity using subtracted spectral density
\begin{eqnarray}
\int_{0}^{\infty} \frac{\Delta \rho(s, \Lambda)}{s-q^2} ds &=& \int_{0}^{\infty} \frac{\Pi^{\rm pert}(s)e^{-s/\Lambda}}{s-q^2}ds + (\pm 2m_s-m_q){\langle \bar
qq\rangle\over q^2}-(\pm 2m_q-m_s){\langle \bar ss\rangle\over
q^2}- {1\over 8}\langle {\alpha_s\over \pi}G^2\rangle{1\over
q^2} \nonumber \\  &\pm & 8\pi\kappa\alpha_s{\langle \bar
qq\rangle\langle\bar ss\rangle\over (q^2)^2}-{16\pi\over
27}\kappa\alpha_s{\langle \bar qq\rangle^2+\langle\bar
ss\rangle^2\over (q^2)^2}. \label{newmatching}
\end{eqnarray}
In the above expression, the dependence on R has been transferred to $\Lambda$. The spectral density is dimensionless. It can be expressed as  $\Delta \rho(s/\Lambda)$. With definition of $x=q^2/\Lambda$ and $y=s/\Lambda$, it is possible to write Eq. (\ref{newmatching}) in the form of
\begin{eqnarray}
\int_{0}^{\infty} \frac{\Delta \rho(s, \Lambda)}{x-y} dy &=& \int_{0}^{\infty} \frac{\Pi^{\rm pert}(s)e^{-y}}{x-y}dy + (\pm 2m_s-m_q){\langle \bar
qq\rangle\over x^2 \Lambda^2}-(\pm 2m_q-m_s){\langle \bar ss\rangle\over
x^2 \Lambda^2}- {1\over 8}\langle {\alpha_s\over \pi}G^2\rangle{1\over
x^2 \Lambda^2} \nonumber \\  &\pm & 8\pi\kappa\alpha_s{\langle \bar
qq\rangle\langle\bar ss\rangle\over (x^2 \Lambda^2)^2}-{16\pi\over
27}\kappa\alpha_s{\langle \bar qq\rangle^2+\langle\bar
ss\rangle^2\over (x^2 \Lambda^2)^2}, \label{newmatching1}
\end{eqnarray}
where $\Lambda$ has been modified to accommodate the condensate terms, ensuring that they are expressed in a dimensionless form. For any finite value of $y$, the quark-hadron duality for the unknown spectral density is not postulated.

It is well-known that, a physical state or resonance should pop up as a peak in the related invariant mass distribution. In this model, the diquark mass should correspond to a peak location of $\Delta\rho(y)$. Therefore, $\Delta\rho(s/\Lambda)$ is a solution for arbitrary $\Lambda$. A physical solution should be insensitive to changes in the $\Lambda$. Consequently, a stability window may emerge when $\Lambda$ increases from a low scale. As the value of $\Lambda$ increases, it becomes vanishingly small in comparison to the condensate contribution in Eq. (\ref{newmatching1}). Subsequently, the solution for $\Delta\rho(y)$ indicates that the peak location of $\Delta\rho(s/\Lambda)$ in $s$ exhibits a dependency on the value of $\Lambda$. This scaling phenomenon makes any $\Delta\rho(y)$ structure impossible to interpret. $\Lambda$ plays an analogous role to the Borel mass parameter in traditional QCDSR.

The inverse problem of QCDSR formalism can be elaborated in the following way. The approach is based on formulating the problem of estimating mass and decay constant as an inverse problem in terms of integral equation. The first kind Fredholm integral equation is expressed as 
\begin{equation}
\int_a^b K(x,t) \varphi(t) dt=f(x), \label{fredint} 
\end{equation}
where the functions $f(x)$ and $K(x,t)$ are given and the function $\varphi(t)$ is the unknown quantity that needs to be determined. In general, an integral equation of this nature represents an inverse problem, wherein a specified kernel $K$ and a driving term $f$ are the fundamental components. In this context, Eq. (\ref{newmatching1}) can be classified as the first kind 
Fredholm integral equation
\begin{equation}
\int_{0}^{\infty} \frac{1}{x-y}\rho(y) dy= \omega(x), \label{fredsr}
\end{equation}
with the unknown function $\rho(y)$ and the input function $\omega(x)$. The objective is to identify an unknown function represented by $\rho(y)$ from an integral equation. Let us suppose that the function $\rho(y)$ decreases at a sufficiently rapid rate with respect to the variable $y$. The integral on the left will be mostly contributed to by a finite range of $y$. This allows the expansion of the integral into a series in terms of $1/x$ to order N for large $\vert x \vert $
\begin{equation}
\frac{1}{x-y}=\sum_{m=1}^N \frac{y^{m-1}}{x^m} \label{expansion1}
\end{equation}
into Eq. (\ref{fredsr}). With the same token, the function $\omega(x)$ can be expanded into a power series in $1/x$  for large $\vert x \vert$
\begin{equation}
\omega(x)=\sum_{n=1}^N \frac{b_n}{x^n}.\label{expansion2}
\end{equation}

The unknown function $\rho(y)$ can be decomposed into 
\begin{equation}
\rho(y)=\sum_{n=1}^N a_ny^\alpha e^{-y}L_{n-1}^{(\alpha)}(y),\label{expansion3}
\end{equation}
in terms of a set of generalized Laguerre functions $L_n^{(\alpha)}$ up to degree $N-1$, which fulfills the orthogonality relation
\begin{equation}
\int_0^\infty y^\alpha e^{-y}L_m^{(\alpha)}(y)L_n^{(\alpha)}(y)dy=\frac{\Gamma(n+\alpha+1)}{n!}\delta_{mn}.
\label{ortlag}
\end{equation} 
The maximal integer $N$ will be determined later. The index $\alpha$ depends on the function $\rho(y)$ behavior near the boundary, where $y$ is close to 0. Upon substituting the values of Eqs. (\ref{expansion1}), (\ref{expansion2}), and (\ref{expansion3}) into Eq. (\ref{fredsr}) and equating the coefficients of $1/x_n$ gives the matrix equation 
\begin{equation}
M a = b, \label{matrixeqn}
\end{equation}
where the matrix elements are
\begin{eqnarray}
M_{mn}=\int_{0}^\infty  y^{m-1+\alpha}e^{-y}L_{n-1}^{(\alpha)}(y) dy,\label{matele}
\end{eqnarray}
with $m$ and $n$ range from 1 to $N$, and the vectors $a=(a_1, a_2,\cdots,a_N)$ and $b=(b_1,b_2,\cdots,b_N)$. Since Eq. (\ref{matrixeqn}) is an equation that involves a matrix, it can be solved via obtaining the inverse matrix, under condition of the existence of inverse matrix. 

If the inverse of $M$ exists, it is possible to obtain a solution for $a$ via the equation $a=M^{-1} b$, with the known input $b$, in a trivial manner. $M^{-1}$ implies a unique solution to $\rho(y)$. Increasing the number of polynomials $N$ provides the optimal solution in Eq. (\ref{expansion3}). A power correction of $1/x^{N+1}$ arises due to the orthogonality observed in Eq. (\ref{ortlag}) between the true and approximate solutions. The orthogonality relation gives $M_{mn}=0$ for $m<n$. This reflects that the matrix $M$ is triangular, and the coefficients $a_n$ previously established remain unaltered when a higher-degree polynomial is incorporated into the expansion  in Eq. (\ref{expansion3}). In a practical application, both $m$ and $n$ must be finite and the determinant of $M$ decreases with its dimension. In such a case, the approximate solution of  $a$ will differ significantly from the true solution when there is a minor fluctuation in the input vector $b$, due to $M^{-1}$. A consequence of severe deviation may be evidenced, for example, by a spectral density losing its positive character. This is a typical ill-posed inverse problem. Accordingly, the optimal $N$ is the integer corresponding to the minimum value of $a_N$ above which a solution is stable or the maximum integer above which a spectral density is positive. The optimal $N$  is set to its maximum value to avoid instability. More details about the method can be found in Ref. \cite{Li:2021gsx} and for the mathematical foundation for this approach, see Ref. \cite{Groetsch:2007}.

\section{Numerical Illustrations}\label{numerical}

The extraction of the masses  and decay constants of $J^P=0^{\pm}$ light-flavor diquarks $(sq)$ and $(ud)$ are obtained with the following set of input \cite{Shifman:1978bx,Ioffe:2005ym,Narison:2005zg,Zhang:2006xp,ParticleDataGroup:2024cfk}:
\begin{eqnarray}
m_s&=& 93.5 \pm 0.8 \ \text{MeV}, \ \langle \bar{q}q  \rangle = (-0.24 \pm 0.01)^3 \ \text{GeV}^3, \  \langle \bar{s}s \rangle=0.8 \langle \bar{q}q  \rangle, \nonumber \\
\langle \frac{\alpha_s}{\pi} G^2 \rangle &=& (0.012 \pm 0.004) \ \text{GeV}^4, \ \alpha_s=0.5, \ \mu \sim 1 \ \text{GeV}, \ \kappa=1.
\end{eqnarray}
The renormalization of $\alpha_s$  and the condensates around 1-2 GeV have a negligible impact on diquark masses. We keep terms with $m_s$ but take $m_s^2 \to 0$. Consequently, these effects are not included in the numerical study. We have also checked that, the deviation from the factorization assumption, where $\kappa=1$ for ideal factorization, did not significantly change the spectroscopic parameters of the diquarks. 

We derive the inverse matrix $M^{-1}$ and unknowns $a_n$  from the OPE expansion coefficients $b_n$ using the boundary conditions $\Delta \rho(y) \sim y$ at $y \to 0$ and $\Delta \rho(y) \to 0$ at $y \to \infty$. The ground state solution is given by the expansion in generalized Laguerre polynomials $\Delta \rho_0(s,\Lambda)=(s/\Lambda)\exp(-s/\Lambda)\sum_{n=1}^N a_n L_{n-1}^{(1)}(s/\Lambda)$. The outcomes of $\Delta \rho_0(s,\Lambda)$ for the scale $\Lambda=5 \ \text{GeV}^2$ with $N=48$ and $N=49$ are effectively indistinguishable, thereby providing assurance that the solutions are stable against variation in $N$. Subsequently, the subtracted spectral densities are solved for using Eq. (\ref{newmatching1}) with a range of values for $\Lambda$. We have checked the $\Lambda$ dependence of diquark masses within $1 \ \text{GeV}^2 \le \Lambda \le 15 \ \text{GeV}^2$ where the masses of $(sq)$ and $(ud)$ diquarks ascends monotonically with $\Lambda$ for $\Lambda> 13 \ \text{GeV}^2$ and are stable wihtin $9 \ \text{GeV}^2 \le \Lambda \le 13 \ \text{GeV}^2$ for $(sq)$ diquark and $5 \ \text{GeV}^2 \le \Lambda \le 9 \ \text{GeV}^2$ for $(ud)$ diquark. As has been argued previously, a physical resonance should be indifferent to the arbitrary scale parameter $\Lambda$. Thus, the stability window $9 \ \text{GeV}^2 \le \Lambda \le 13 \ \text{GeV}^2$ for $(sq)$ diquark and $5 \ \text{GeV}^2 \le \Lambda \le 9 \ \text{GeV}^2$ for $(ud)$ diquark lead to the masses as
\begin{eqnarray}
M_{sq}(0^+)=847 \pm 2 \ \text{MeV}, \ M_{sq}(0^-)=845 \pm 2 \ \text{MeV}, \nonumber \\
M_{ud}(0^+)=678 \pm 3 \ \text{MeV}, \ M_{ud}(0^-)=672 \pm 3 \ \text{MeV}.
\end{eqnarray}

\begin{figure}[h]
\begin{center}
\includegraphics[totalheight=6cm,width=8cm]{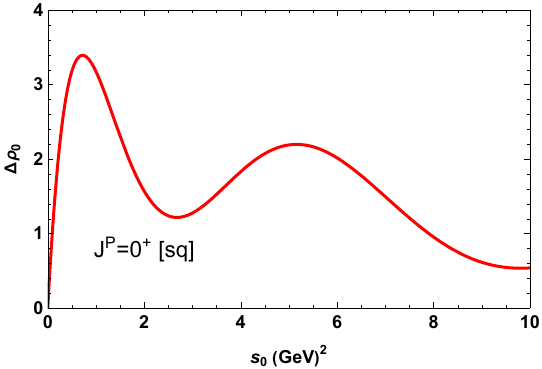}
\includegraphics[totalheight=6cm,width=8cm]{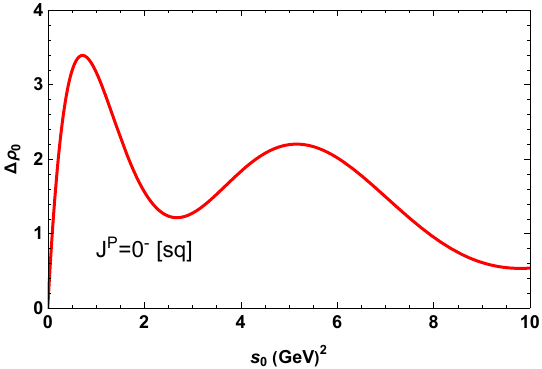}
\end{center}
\caption{$s$ dependence of the ground state solution $\Delta \rho_0(s,\Lambda)$ for 
$\Lambda=11.0 \ \text{GeV}^2$ for $J^P=0^+$ [sq] diquark (left panel), and  for $J^P=0^-$ [sq] diquark (right panel).}
\label{spectralsq}
\end{figure}

\begin{figure}[h]
\begin{center}
\includegraphics[totalheight=6cm,width=8cm]{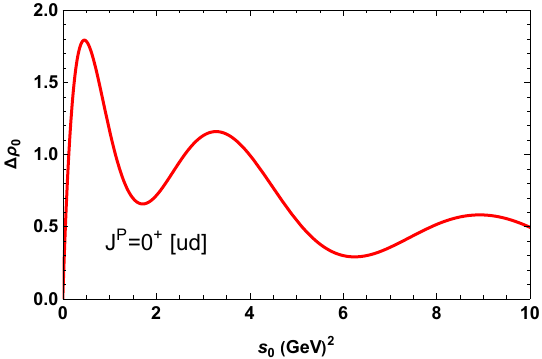}
\includegraphics[totalheight=6cm,width=8cm]{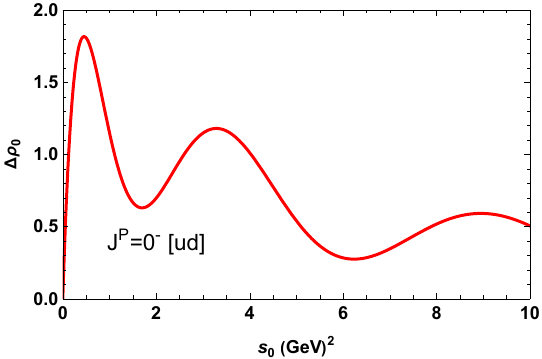}
\end{center}
\caption{$s$ dependence of the ground state solution $\Delta \rho_0(s,\Lambda)$ for 
$\Lambda=7.0\ \text{GeV}^2$ for $J^P=0^+$ [ud] diquark  (left panel), and for  $J^P=0^-$ [ud] diquark (right panel).}
\label{spectralud}
\end{figure}

The uncertainties are attributed to two primary sources: firstly, the values of all input parameters are subject to error, and secondly, the calculations of the stability window are also susceptible to inaccuracy. The uncertainties of the extracted masses are well below within the accuracy of the standard technique. We observe that, the masses of $J^P=0^+$ and $J^P=0^-$ are almost degenerate. These features can be traced down in Figs. \ref{spectralsq} and \ref{spectralud} where we plot the subtracted spectral densities for $(sq)$ and $(ud)$ diquarks, respectively. 

We also compare our results with those available in the literature which can be seen in Table \ref{masscompare}. Refs. \cite{Ebert:1995fp,Ebert:1996ab,Ebert:2007nw} used a relativistic constituent quark model, Ref. \cite{Burden:1996nh} solved two body Salpeter approach, Ref.  \cite{Hess:1998sd} solved a Dyson-Schwinger equation, Ref. \cite{Maris:2002yu} used a lattice QCD approach and Ref. \cite{Wang:2011ab} applied QCDSR approach. Our result of $(sq)$ diquark agree well with the results of Refs. \cite{Ebert:1995fp,Ebert:1996ab,Burden:1996nh}. The mass value for the $(sq)$ diquark is approximately 100 MeV lower than that reported in Ref. \cite{Ebert:2007nw}, 250 MeV lower than  Ref. \cite{Hess:1998sd}, and 70 MeV higher than Ref.\cite{Wang:2011ab}. For the $ud$ diquark, our result agree well with the predictions of Refs. \cite{Ebert:2007nw,Ebert:1995fp,Ebert:1996ab,Maris:2002yu,Wang:2011ab} whereas it is approximately 150 MeV lower than Ref. \cite{Hess:1998sd}. It is important to note that some of the observed discrepancies in the results may be attributed to underlying assumptions in the methodologies employed, as well as differences in the nature of interactions within the bound states and in the parameters utilized as inputs. 

\begin{table}[h!]
\caption{Extracted masses of $sq$ and $ud$ diquarks with $J^P=0^+$ quantum number. The results are presented in unit of MeV.} \label{masscompare}
\begin{tabular}{c|c|c|c|c|c|c|c}
\toprule
State  &  This work &   \cite{Ebert:2007nw} & \cite{Ebert:1995fp}, \cite{Ebert:1996ab}&\cite{Burden:1996nh} & \cite{Hess:1998sd}& \cite{Maris:2002yu} &\cite{Wang:2011ab}  \\
\toprule
$sq$  &  $847 \pm 2$  & 948 &  895 & 882 & 1100 &- & $0.77 \pm  0.04$\\
$ud$  &  $678 \pm 3$  &  710  &  705  & 737 & 820 &  694(22) & $0.64 \pm  0.06$ \\
\toprule
\end{tabular}
\end{table}

The decay constants in this formalism can be obtained as follows. In reference to the aforementioned literature, as demonstrated in Ref. \cite{CP-PACS:2001ncr}, the area under the resonance peak of $\rho(s)$, is equivalent to the square of the decay constant, $f$. In this formulation, the resonance peak is accurately represented by the subtracted spectral density $ \Delta \rho(s)$ which has largely removed the continuum contribution and decay constant can be formulated as
\begin{equation}
f^2 \thickapprox \int_0^\infty \Delta \rho(s) ds. \label{decayconstant}
\end{equation}
Using this formula, we list our decay constant values together with the results available in the literature in Table \ref{decaycompare}. We have averaged the decay constants of Ref. \cite{Jamin:1989hh} for the related spin-parity quantum number. As is clear in table, the results are compatible with the references in which QCDSR formalism is applied. 

\begin{table}[h!]
\caption{Extracted decay constants $f_d$ of $sq$ and $ud$ diquarks with $J^P=0^{\pm}$ quantum numbers. Results are presented in unit of $\text{GeV}^2$.} \label{decaycompare}
\begin{tabular}{c|c|c|c|c}
\toprule
State  & $J^P$ &This work & \cite{Jamin:1989hh}  &\cite{Wang:2011ab}  \\
\toprule
$sq$  &  $0^+$ & $0.218 \pm  0.005 \ $ & $0.219 \pm 0.019$ & $0.313 \pm  0.013 $ \\
$sq$  &  $0^-$ & $0.217 \pm  0.004 $ & $0.21\pm 0.005$ &- \\
$ud$  &  $0^+$ & $0.209 \pm  0.005 $ & $0.184 \pm 0.023$ &$0.264 \pm  0.017 $ \\
$ud$  &  $0^-$ & $0.208 \pm  0.005 $ &$0.21 \pm 0.005$  & - \\
\toprule
\end{tabular}
\end{table}

\section{Epilogue}\label{final}

QCDSR is founded on two key principles: the analyticity of the two-point correlation function and the asymptotic freedom of QCD. The first principle enables the derivation of dispersion relations and the second principle, asymptotic freedom, allows for the systematic computation of the correlation function in the deep Euclidean region through the OPE. The OPE is linked to the growth of nonperturbative operator values and the associated Wilson coefficients, which describe short-distance correlator dynamics. These coefficients can be determined using perturbative methods. As a result, equations relating specific integrals of the spectral function—sums of the contributions of physical states—to the outcome of the OPE are obtained. This is where "sum rules" come into play \cite{Gubler:2018ctz}. Moreover, the high-energy component of the spectral function is substituted with the OPE expression, which is analytically derived based on the principle of quark-hadron duality. Quark-hadron duality, first proposed in Ref. \cite{Poggio:1975af}, posits that an experimentally measurable hadronic spectral function $\rho(s)$ averaged over a specified energy range, can be described by the corresponding expression calculated from  QCD using its fundamental degrees of freedom, namely quarks and gluons. This duality is essential for understanding the transition between the perturbative and nonperturbative regimes of QCD. Integrals that encompass only the low-energy portion of the spectral function can be derived from QCD via the application of the OPE.

The QCDSR framework is a widely used approach for addressing nonperturbative QCD, yielding predictions that often align with experimental observations. When a state is initially observed, a QCDSR calculation can be instrumental in determining its existence and characterizing its properties. Such calculations can provide evidence either supporting or refuting the existence of the state in question.

Although the method was proposed for mesons, it has also yielded results consistent with experimental observations for situations involving multiquark states. As such, the method is very useful for nonperturbative QCD and is one of the few methods that can give analytical solutions. Some auxiliary parameters (continuum threshold and Borel mass parameters) arising from the method affect the margins of error in the results predicted by the method. Therefore, some new approaches have recently been proposed to improve the predicted results of the QCDSR method, as mentioned in the formalism. The inverse matrix method is one of them.

In the inverse matrix method, the unknown hadron spectral density is represented as an expansion in terms of orthogonal basis functions. The coefficients of $1/(q^2)^m$  are equated to convert the integral equation into a matrix equation.  The solution to the matrix equation will yield potential solutions for the spectral density. The inverse problem approach to QCDSR provides a robust and simplified methodology for the investigation of nonperturbative QCD phenomena. By directly solving for the spectral density without assuming quark-hadron duality, this method provides accurate determinations of resonance properties and decay widths, with broad applications expected in low-energy QCD observables. 

Using the inverse matrix method for QCDSR approach, we revisited the spectroscopic parameters of light-flavor diquark states. Diquarks are particles with a color charge formed by two quarks coming together. Although no diquarks have been observed experimentally to date, they are thought to come together to form tetraquarks, pentaquarks and hexaquarks. The mass and decomposition constant values obtained are compared with the results obtained by different methods in the existing literature. The results obtained in this study are in accordance with those reported in the existing literature. 

It is anticipated that the results will contribute to a deeper comprehension of the interactions and internal structure of exotic particles. Within this framework, the accuracy of theoretical predictions can be systematically refined by incorporating higher-order and higher-power corrections on the quark side.

\bibliography{inversediquark-revised}

\begin{thebibliography}{10}
\expandafter\ifx\csname url\endcsname\relax
  \def\url#1{\texttt{#1}}\fi
\expandafter\ifx\csname urlprefix\endcsname\relax\def\urlprefix{URL }\fi
\expandafter\ifx\csname href\endcsname\relax
  \def\href#1#2{#2} \def\path#1{#1}\fi

\bibitem{Shifman:1978bx}
M.~A. Shifman, A.~I. Vainshtein, V.~I. Zakharov, {QCD and Resonance Physics.
  Theoretical Foundations}, Nucl. Phys. B 147 (1979) 385--447.
\newblock \href {https://doi.org/10.1016/0550-3213(79)90022-1}
  {\path{doi:10.1016/0550-3213(79)90022-1}}.

\bibitem{Shifman:1978by}
M.~A. Shifman, A.~I. Vainshtein, V.~I. Zakharov, {QCD and Resonance Physics:
  Applications}, Nucl. Phys. B 147 (1979) 448--518.
\newblock \href {https://doi.org/10.1016/0550-3213(79)90023-3}
  {\path{doi:10.1016/0550-3213(79)90023-3}}.

\bibitem{Cohen:1994wm}
T.~D. Cohen, R.~J. Furnstahl, D.~K. Griegel, X.-m. Jin, {QCD sum rules and
  applications to nuclear physics}, Prog. Part. Nucl. Phys. 35 (1995) 221--298.
\newblock \href {http://arxiv.org/abs/hep-ph/9503315}
  {\path{arXiv:hep-ph/9503315}}, \href
  {https://doi.org/10.1016/0146-6410(95)00043-I}
  {\path{doi:10.1016/0146-6410(95)00043-I}}.

\bibitem{Nielsen:2009uh}
M.~Nielsen, F.~S. Navarra, S.~H. Lee, {New Charmonium States in QCD Sum Rules:
  A Concise Review}, Phys. Rept. 497 (2010) 41--83.
\newblock \href {http://arxiv.org/abs/0911.1958} {\path{arXiv:0911.1958}},
  \href {https://doi.org/10.1016/j.physrep.2010.07.005}
  {\path{doi:10.1016/j.physrep.2010.07.005}}.

\bibitem{Albuquerque:2018jkn}
R.~M. Albuquerque, J.~M. Dias, K.~P. Khemchandani, A.~Mart\'\i{}nez~Torres,
  F.~S. Navarra, M.~Nielsen, C.~M. Zanetti, {QCD sum rules approach to the
  $X,~Y$ and $Z$ states}, J. Phys. G 46~(9) (2019) 093002.
\newblock \href {http://arxiv.org/abs/1812.08207} {\path{arXiv:1812.08207}},
  \href {https://doi.org/10.1088/1361-6471/ab2678}
  {\path{doi:10.1088/1361-6471/ab2678}}.

\bibitem{Narison:1989aq}
S.~Narison, {QCD spectral sum rules}, Vol.~26, 1989.

\bibitem{Narison:2002woh}
S.~Narison, {QCD as a Theory of Hadrons : From Partons to Confinement},
  Vol.~17, Oxford University Press, 2005.
\newblock \href {http://arxiv.org/abs/hep-ph/0205006}
  {\path{arXiv:hep-ph/0205006}}, \href {https://doi.org/10.1017/9781009290296}
  {\path{doi:10.1017/9781009290296}}.

\bibitem{Alexandrou:2006cq}
C.~Alexandrou, P.~de~Forcrand, B.~Lucini, {Evidence for diquarks in lattice
  QCD}, Phys. Rev. Lett. 97 (2006) 222002.
\newblock \href {http://arxiv.org/abs/hep-lat/0609004}
  {\path{arXiv:hep-lat/0609004}}, \href
  {https://doi.org/10.1103/PhysRevLett.97.222002}
  {\path{doi:10.1103/PhysRevLett.97.222002}}.

\bibitem{Gell-Mann:1964ewy}
M.~Gell-Mann, {A Schematic Model of Baryons and Mesons}, Phys. Lett. 8 (1964)
  214--215.
\newblock \href {https://doi.org/10.1016/S0031-9163(64)92001-3}
  {\path{doi:10.1016/S0031-9163(64)92001-3}}.

\bibitem{Zweig:1964jf}
G.~Zweig, {An SU(3) model for strong interaction symmetry and its breaking.
  Version 2}, 1964, pp. 22--101.
\newblock \href {https://doi.org/10.17181/CERN-TH-412}
  {\path{doi:10.17181/CERN-TH-412}}.

\bibitem{Jaffe:2003sg}
R.~L. Jaffe, F.~Wilczek, {Diquarks and exotic spectroscopy}, Phys. Rev. Lett.
  91 (2003) 232003.
\newblock \href {http://arxiv.org/abs/hep-ph/0307341}
  {\path{arXiv:hep-ph/0307341}}, \href
  {https://doi.org/10.1103/PhysRevLett.91.232003}
  {\path{doi:10.1103/PhysRevLett.91.232003}}.

\bibitem{Jaffe:2004ph}
R.~L. Jaffe, {Exotica}, Phys. Rept. 409 (2005) 1--45.
\newblock \href {http://arxiv.org/abs/hep-ph/0409065}
  {\path{arXiv:hep-ph/0409065}}, \href
  {https://doi.org/10.1016/j.physrep.2004.11.005}
  {\path{doi:10.1016/j.physrep.2004.11.005}}.

\bibitem{Wilczek:2004im}
F.~Wilczek, {Diquarks as inspiration and as objects}, in: {Deserfest: A
  Celebration of the Life and Works of Stanley Deser}, 2004, pp. 322--338.
\newblock \href {http://arxiv.org/abs/hep-ph/0409168}
  {\path{arXiv:hep-ph/0409168}}, \href
  {https://doi.org/10.1142/9789812775344_0007}
  {\path{doi:10.1142/9789812775344_0007}}.

\bibitem{Selem:2006nd}
A.~Selem, F.~Wilczek, {Hadron systematics and emergent diquarks}, in: {Ringberg
  Workshop on New Trends in HERA Physics 2005}, 2006, pp. 337--356.
\newblock \href {http://arxiv.org/abs/hep-ph/0602128}
  {\path{arXiv:hep-ph/0602128}}, \href
  {https://doi.org/10.1142/9789812773524_0030}
  {\path{doi:10.1142/9789812773524_0030}}.

\bibitem{Barabanov:2020jvn}
M.~Y. Barabanov, et~al., {Diquark correlations in hadron physics: Origin,
  impact and evidence}, Prog. Part. Nucl. Phys. 116 (2021) 103835.
\newblock \href {http://arxiv.org/abs/2008.07630} {\path{arXiv:2008.07630}},
  \href {https://doi.org/10.1016/j.ppnp.2020.103835}
  {\path{doi:10.1016/j.ppnp.2020.103835}}.

\bibitem{Leinweber:1995fn}
D.~B. Leinweber, {QCD sum rules for skeptics}, Annals Phys. 254 (1997)
  328--396.
\newblock \href {http://arxiv.org/abs/nucl-th/9510051}
  {\path{arXiv:nucl-th/9510051}}, \href
  {https://doi.org/10.1006/aphy.1996.5641} {\path{doi:10.1006/aphy.1996.5641}}.

\bibitem{Carvunis:2024koh}
A.~Carvunis, F.~Mahmoudi, Y.~Monceaux, {Potential of light-cone sum rules
  without semiglobal quark-hadron duality}, Phys. Rev. D 110~(11) (2024)
  114008.
\newblock \href {http://arxiv.org/abs/2404.01290} {\path{arXiv:2404.01290}},
  \href {https://doi.org/10.1103/PhysRevD.110.114008}
  {\path{doi:10.1103/PhysRevD.110.114008}}.

\bibitem{Li:2020xrz}
H.-N. Li, H.~Umeeda, F.~Xu, F.-S. Yu, {$D$ meson mixing as an inverse problem},
  Phys. Lett. B 810 (2020) 135802.
\newblock \href {http://arxiv.org/abs/2001.04079} {\path{arXiv:2001.04079}},
  \href {https://doi.org/10.1016/j.physletb.2020.135802}
  {\path{doi:10.1016/j.physletb.2020.135802}}.

\bibitem{Li:2020ejs}
H.-n. Li, H.~Umeeda, {QCD sum rules with spectral densities solved in inverse
  problems}, Phys. Rev. D 102 (2020) 114014.
\newblock \href {http://arxiv.org/abs/2006.16593} {\path{arXiv:2006.16593}},
  \href {https://doi.org/10.1103/PhysRevD.102.114014}
  {\path{doi:10.1103/PhysRevD.102.114014}}.

\bibitem{Xiong:2022uwj}
A.-S. Xiong, T.~Wei, F.-S. Yu, {Inverse Problem Approach for Non-Perturbative
  QCD: Foundation} (11 2022).
\newblock \href {http://arxiv.org/abs/2211.13753} {\path{arXiv:2211.13753}}.

\bibitem{Li:2021gsx}
H.-n. Li, {Dispersive analysis of glueball masses}, Phys. Rev. D 104~(11)
  (2021) 114017.
\newblock \href {http://arxiv.org/abs/2109.04956} {\path{arXiv:2109.04956}},
  \href {https://doi.org/10.1103/PhysRevD.104.114017}
  {\path{doi:10.1103/PhysRevD.104.114017}}.

\bibitem{Li:2022qul}
H.-n. Li, {Dispersive derivation of the pion distribution amplitude}, Phys.
  Rev. D 106~(3) (2022) 034015.
\newblock \href {http://arxiv.org/abs/2205.06746} {\path{arXiv:2205.06746}},
  \href {https://doi.org/10.1103/PhysRevD.106.034015}
  {\path{doi:10.1103/PhysRevD.106.034015}}.

\bibitem{Li:2022jxc}
H.-n. Li, {Dispersive analysis of neutral meson mixing}, Phys. Rev. D 107~(5)
  (2023) 054023.
\newblock \href {http://arxiv.org/abs/2208.14798} {\path{arXiv:2208.14798}},
  \href {https://doi.org/10.1103/PhysRevD.107.054023}
  {\path{doi:10.1103/PhysRevD.107.054023}}.

\bibitem{Li:2023dqi}
H.-n. Li, {Dispersive constraints on fermion masses}, Phys. Rev. D 107~(9)
  (2023) 094007.
\newblock \href {http://arxiv.org/abs/2302.01761} {\path{arXiv:2302.01761}},
  \href {https://doi.org/10.1103/PhysRevD.107.094007}
  {\path{doi:10.1103/PhysRevD.107.094007}}.

\bibitem{Li:2023yay}
H.-n. Li, {Dispersive determination of electroweak-scale masses}, Phys. Rev. D
  108~(5) (2023) 054020.
\newblock \href {http://arxiv.org/abs/2304.05921} {\path{arXiv:2304.05921}},
  \href {https://doi.org/10.1103/PhysRevD.108.054020}
  {\path{doi:10.1103/PhysRevD.108.054020}}.

\bibitem{Li:2023ncg}
H.-n. Li, {Dispersive determination of neutrino mass ordering} (6 2023).
\newblock \href {http://arxiv.org/abs/2306.03463} {\path{arXiv:2306.03463}}.

\bibitem{Li:2023fim}
H.-n. Li, {Dispersive determination of fourth generation quark masses}, Phys.
  Rev. D 109~(11) (2024) 115024.
\newblock \href {http://arxiv.org/abs/2309.15602} {\path{arXiv:2309.15602}},
  \href {https://doi.org/10.1103/PhysRevD.109.115024}
  {\path{doi:10.1103/PhysRevD.109.115024}}.

\bibitem{Li:2024awx}
H.-n. Li, {Understanding small neutrino mass and its implication}, Chin. J.
  Phys. 92 (2024) 1043--1054.
\newblock \href {http://arxiv.org/abs/2404.16626} {\path{arXiv:2404.16626}},
  \href {https://doi.org/10.1016/j.cjph.2024.10.009}
  {\path{doi:10.1016/j.cjph.2024.10.009}}.

\bibitem{Li:2024xnl}
H.-n. Li, {Dispersive determination of the fourth generation lepton masses}, J.
  Phys. G 52~(2) (2025) 025001.
\newblock \href {http://arxiv.org/abs/2407.07813} {\path{arXiv:2407.07813}},
  \href {https://doi.org/10.1088/1361-6471/ada0cd}
  {\path{doi:10.1088/1361-6471/ada0cd}}.

\bibitem{Zhao:2024drr}
Z.-X. Zhao, Y.-P. Xing, R.-H. Li, {A progress in the inverse matrix method in
  QCD sum rules}, Eur. Phys. J. C 84~(10) (2024) 1105.
\newblock \href {http://arxiv.org/abs/2407.09819} {\path{arXiv:2407.09819}},
  \href {https://doi.org/10.1140/epjc/s10052-024-13452-8}
  {\path{doi:10.1140/epjc/s10052-024-13452-8}}.

\bibitem{Li:2024fko}
H.-n. Li, {Dispersive Analysis of Excited Glueball States}, Chin. Phys. Lett.
  41~(10) (2024) 101101.
\newblock \href {http://arxiv.org/abs/2408.06738} {\path{arXiv:2408.06738}},
  \href {https://doi.org/10.1088/0256-307X/41/10/101101}
  {\path{doi:10.1088/0256-307X/41/10/101101}}.

\bibitem{Dosch:1988hu}
H.~G. Dosch, M.~Jamin, B.~Stech, {Diquarks, {QCD} Sum Rules and Weak Decays},
  Z. Phys. C 42 (1989) 167.
\newblock \href {https://doi.org/10.1007/BF01565139}
  {\path{doi:10.1007/BF01565139}}.

\bibitem{Jamin:1989hh}
M.~Jamin, M.~Neubert, {Diquark Decay Constants From {QCD} Sum Rules}, Phys.
  Lett. B 238 (1990) 387--394.
\newblock \href {https://doi.org/10.1016/0370-2693(90)91753-X}
  {\path{doi:10.1016/0370-2693(90)91753-X}}.

\bibitem{Groetsch:2007}
W.~C. Groetsch, {Integral equations of the first kind, inverse problems and
  regularization: a crash course}, J. Phys.: Conf. Ser. 73 (2007) 012001.
\newblock \href {https://doi.org/10.1088/1742-6596/73/1/012001}
  {\path{doi:10.1088/1742-6596/73/1/012001}}.

\bibitem{Ioffe:2005ym}
B.~L. Ioffe, {QCD at low energies}, Prog. Part. Nucl. Phys. 56 (2006) 232--277.
\newblock \href {http://arxiv.org/abs/hep-ph/0502148}
  {\path{arXiv:hep-ph/0502148}}, \href
  {https://doi.org/10.1016/j.ppnp.2005.05.001}
  {\path{doi:10.1016/j.ppnp.2005.05.001}}.

\bibitem{Narison:2005zg}
S.~Narison, {Strange quark, tachyonic gluon masses and V(us) from hadronic tau
  decays}, Phys. Lett. B 626 (2005) 101--110.
\newblock \href {http://arxiv.org/abs/hep-ph/0501208}
  {\path{arXiv:hep-ph/0501208}}, \href
  {https://doi.org/10.1016/j.physletb.2005.08.085}
  {\path{doi:10.1016/j.physletb.2005.08.085}}.

\bibitem{Zhang:2006xp}
A.~Zhang, T.~Huang, T.~G. Steele, {Diquark and light four-quark states}, Phys.
  Rev. D 76 (2007) 036004.
\newblock \href {http://arxiv.org/abs/hep-ph/0612146}
  {\path{arXiv:hep-ph/0612146}}, \href
  {https://doi.org/10.1103/PhysRevD.76.036004}
  {\path{doi:10.1103/PhysRevD.76.036004}}.

\bibitem{ParticleDataGroup:2024cfk}
S.~Navas, et~al., {Review of particle physics}, Phys. Rev. D 110~(3) (2024)
  030001.
\newblock \href {https://doi.org/10.1103/PhysRevD.110.030001}
  {\path{doi:10.1103/PhysRevD.110.030001}}.

\bibitem{Ebert:1995fp}
D.~Ebert, T.~Feldmann, C.~Kettner, H.~Reinhardt, {A Diquark model for baryons
  containing one heavy quark}, Z. Phys. C 71 (1996) 329--336.
\newblock \href {http://arxiv.org/abs/hep-ph/9506298}
  {\path{arXiv:hep-ph/9506298}}, \href {https://doi.org/10.1007/BF02906991}
  {\path{doi:10.1007/BF02906991}}.

\bibitem{Ebert:1996ab}
D.~Ebert, T.~Feldmann, C.~Kettner, H.~Reinhardt, {Heavy baryons in the
  quark-diquark picture}, Int. J. Mod. Phys. A 13 (1998) 1091--1113.
\newblock \href {http://arxiv.org/abs/hep-ph/9601257}
  {\path{arXiv:hep-ph/9601257}}, \href
  {https://doi.org/10.1142/S0217751X98000482}
  {\path{doi:10.1142/S0217751X98000482}}.

\bibitem{Ebert:2007nw}
D.~Ebert, R.~N. Faustov, V.~O. Galkin, {Masses of excited heavy baryons in the
  relativistic quark model}, Phys. Lett. B 659 (2008) 612--620.
\newblock \href {http://arxiv.org/abs/0705.2957} {\path{arXiv:0705.2957}},
  \href {https://doi.org/10.1016/j.physletb.2007.11.037}
  {\path{doi:10.1016/j.physletb.2007.11.037}}.

\bibitem{Burden:1996nh}
C.~J. Burden, L.~Qian, C.~D. Roberts, P.~C. Tandy, M.~J. Thomson, {Ground state
  spectrum of light quark mesons}, Phys. Rev. C 55 (1997) 2649--2664.
\newblock \href {http://arxiv.org/abs/nucl-th/9605027}
  {\path{arXiv:nucl-th/9605027}}, \href
  {https://doi.org/10.1103/PhysRevC.55.2649}
  {\path{doi:10.1103/PhysRevC.55.2649}}.

\bibitem{Hess:1998sd}
M.~Hess, F.~Karsch, E.~Laermann, I.~Wetzorke, {Diquark masses from lattice
  QCD}, Phys. Rev. D 58 (1998) 111502.
\newblock \href {http://arxiv.org/abs/hep-lat/9804023}
  {\path{arXiv:hep-lat/9804023}}, \href
  {https://doi.org/10.1103/PhysRevD.58.111502}
  {\path{doi:10.1103/PhysRevD.58.111502}}.

\bibitem{Maris:2002yu}
P.~Maris, {Effective masses of diquarks}, Few Body Syst. 32 (2002) 41--52.
\newblock \href {http://arxiv.org/abs/nucl-th/0204020}
  {\path{arXiv:nucl-th/0204020}}, \href
  {https://doi.org/10.1007/s00601-002-0111-7}
  {\path{doi:10.1007/s00601-002-0111-7}}.

\bibitem{Wang:2011ab}
Z.-G. Wang, {Analysis of the light-flavor scalar and axial-vector diquark
  states with QCD sum rules}, Commun. Theor. Phys. 59 (2013) 451--456.
\newblock \href {http://arxiv.org/abs/1112.5910} {\path{arXiv:1112.5910}},
  \href {https://doi.org/10.1088/0253-6102/59/4/11}
  {\path{doi:10.1088/0253-6102/59/4/11}}.

\bibitem{CP-PACS:2001ncr}
T.~Yamazaki, et~al., {Spectral function and excited states in lattice QCD with
  maximum entropy method}, Phys. Rev. D 65 (2002) 014501.
\newblock \href {http://arxiv.org/abs/hep-lat/0105030}
  {\path{arXiv:hep-lat/0105030}}, \href
  {https://doi.org/10.1103/PhysRevD.65.014501}
  {\path{doi:10.1103/PhysRevD.65.014501}}.

\bibitem{Gubler:2018ctz}
P.~Gubler, D.~Satow, {Recent Progress in QCD Condensate Evaluations and Sum
  Rules}, Prog. Part. Nucl. Phys. 106 (2019) 1--67.
\newblock \href {http://arxiv.org/abs/1812.00385} {\path{arXiv:1812.00385}},
  \href {https://doi.org/10.1016/j.ppnp.2019.02.005}
  {\path{doi:10.1016/j.ppnp.2019.02.005}}.

\bibitem{Poggio:1975af}
E.~C. Poggio, H.~R. Quinn, S.~Weinberg, {Smearing the Quark Model}, Phys. Rev.
  D 13 (1976) 1958.
\newblock \href {https://doi.org/10.1103/PhysRevD.13.1958}
  {\path{doi:10.1103/PhysRevD.13.1958}}.

\end{thebibliography}

\end{document}